\documentclass[twocolumn,groupedaddress,showpacs, superscriptaddress,prb]{revtex4-1} 

\usepackage{amsmath}
\usepackage{amssymb}
\usepackage{gensymb}
\usepackage{graphicx,bm,units,yfonts}
\usepackage[table]{xcolor}
\usepackage{hyperref}
\usepackage{movie15}
\usepackage{color}

\newcommand{\RM}{{\mathbb R}}

\newcommand{\TM}{{\mathbb T}}
\newcommand{\ZM}{{\mathbb Z}}

\newcommand{\Tt}{{\mathcal T}}

\newcommand{\Pp}{{\mathcal P}}
\newcommand{\Ss}{{\mathcal S}}
\newcommand{\Ll}{{\mathcal L}}

\begin{document}

\title{Revealing the boundary Weyl physics of the four-dimensional Hall effect via phason engineering in metamaterials}

\author{Wenting Cheng}
\affiliation{Department of Physics, New Jersey Institute of Technology, Newark, NJ, USA}

\author{Emil Prodan}
\email{prodan@yu.edu}
\affiliation{Department of Physics, Yeshiva University, New York, NY, USA}

\author{Camelia Prodan}
\email{cprodan@njit.edu}
\affiliation{Department of Physics, New Jersey Institute of Technology, Newark, NJ, USA}

\begin{abstract}
Quantum Hall physics has been theoretically predicted in 4-dimensions and higher. In hypothetical 2n-dimensions, the topological characters of both the bulk and the boundary are manifested as quantized non-linear transport coefficients that connect, respectively, to the n-th Chern number of the bulk gap projection and to the n-th winding number of the Weyl spectral singularities on the (2n-1)-dimensional boundaries. Here, we introduce the concept of phason engineering in metamaterials and use it as a vehicle to access and apply the quantum Hall physics in arbitrary dimensions. Using these specialized design principles, we fabricate a re-configurable 2-dimensional aperiodic acoustic crystal with a phason living on a 2-torus, giving us access to the 4-dimensional quantum Hall physics. Also, we supply a direct experimental confirmation that the topological boundary spectrum assembles in a Weyl singularity when mapped as function of the quasi-momenta. We also demonstrate topological wave steering enabled by the Weyl physics of the 3-dimensional boundaries.
\end{abstract}

\maketitle

\section{Introduction}
 In 1988, Haldane predicted that quantized Hall physics can be intrinsic to a material \cite{haldane1988model}. To generate the effect without an external magnetic field, he had to consider an atomic lattice with at least two degrees of freedom per repeating cell and to rely on complex hopping parameters that break the time-reversal symmetry. Chromium-doped thin films of (Bi,Sb)$_2$Te$_3$ produced the experimental validation of Haldane's prediction \cite{ChangAdvMat2013,ChangScience2013}. Two decades after Haldane's seminal work, the quantum Hall physics was predicted to also manifest in systems with electro-magnetic \cite{HaldanePRL2008} and mechanical \cite{ProdanPRL2009} degrees of freedom. These predictions became reality in 2009, when confirmed with gyromagnetic photonic crystals \cite{WangNature2009}, and in 2015, when confirmed with gyroscope lattices \cite{NashPNAS2015}, respectively. These earlier advances flourished in extremely active fields, where quantum Hall physics is investigated with both quantum materials and classical meta-materials.

Integer quantum Hall effect (IQHE) generalizes in 4-dimensions (4D) and higher,\cite{ZhangScience2001} and the representative theoretical models that display the effect, intrinsically, have been already enumerated \cite{RyuNJP2010} [see also \cite[Sec.~2.2.4]{ProdanSpringer2016}]. Those representative models assume spatial periodicity and this simplified setting comes at the price of increased complexity in the degrees of freedom per repeating cell. For example, the simplest model in dimension $d=2n$ requires $2^n$ degrees of freedom, complex connectivity and a high level of tuning to stabilize a topological phase \cite{RyuNJP2010}. One strategy for implementing such higher dimensional models is to see them as supplying labels for and specific connections between the degrees of freedom. As long as these labels and their connections are identically reproduced, the degrees of freedom can be rendered in any dimension, in particular, in our 3-dimensional physical space. Following this strategy, the 4D QHE was recently implemented with classical electric circuits \cite{WangNaatComm2020}. While certainly  an impressive demonstration, the outcome was an extremely complex network of connected circuit components.

Starting with the work of Kraus {\it et al},\cite{KrausPRL2012} a strategy emerged for the emulation of topological effects from higher dimensions. It relies on aperiodicity, specifically, on the fact that any aperiodic pattern has an intrinsic degree of freedom, the phason, which, at least in principle, can be engineered, accessed and controlled experimentally.\cite{ApigoPRM2018,ProdanJGP2019} The phason space augments the physical space and this opens a door to higher dimensional physics. Certainly, the experimental emulations of the 4D QHE were based on these principles.\cite{LohseNature2018,ZilberbergNature2018} Working with ultracold atoms, Lohse {\it et al} \cite{LohseNature2018} were able to map a cloud's center of mass as it navigated an aperiodically modulated potential and to demonstrate the quantization of the bulk topological invariant via a connection established in an earlier theoretical work.\cite{PricePRL2015} The bulk-boundary correspondence was not addressed in this study. Zilberberg {\it et al}\cite{ZilberbergNature2018} emulated the 4D QHE with spatially modulated arrays of coupled optical wave guides and produced evidence of topological boundary modes. However, due to the specific physics involved, the analysis rested entirely on the spatial profile of the modes and their actual energies were not resolved. As such, no evidence of the hallmark Weyl singularity in the dispersion of the boundary modes was presented, predicted by the {\it strong} bulk-boundary correspondence of 4D QHE \cite{RyuNJP2010}. Instead, other weaker forms of bulk-boundary correspondence were presented, such as the corner-to-corner pumping. 

\begin{figure*}[t]
\includegraphics[width=\linewidth]{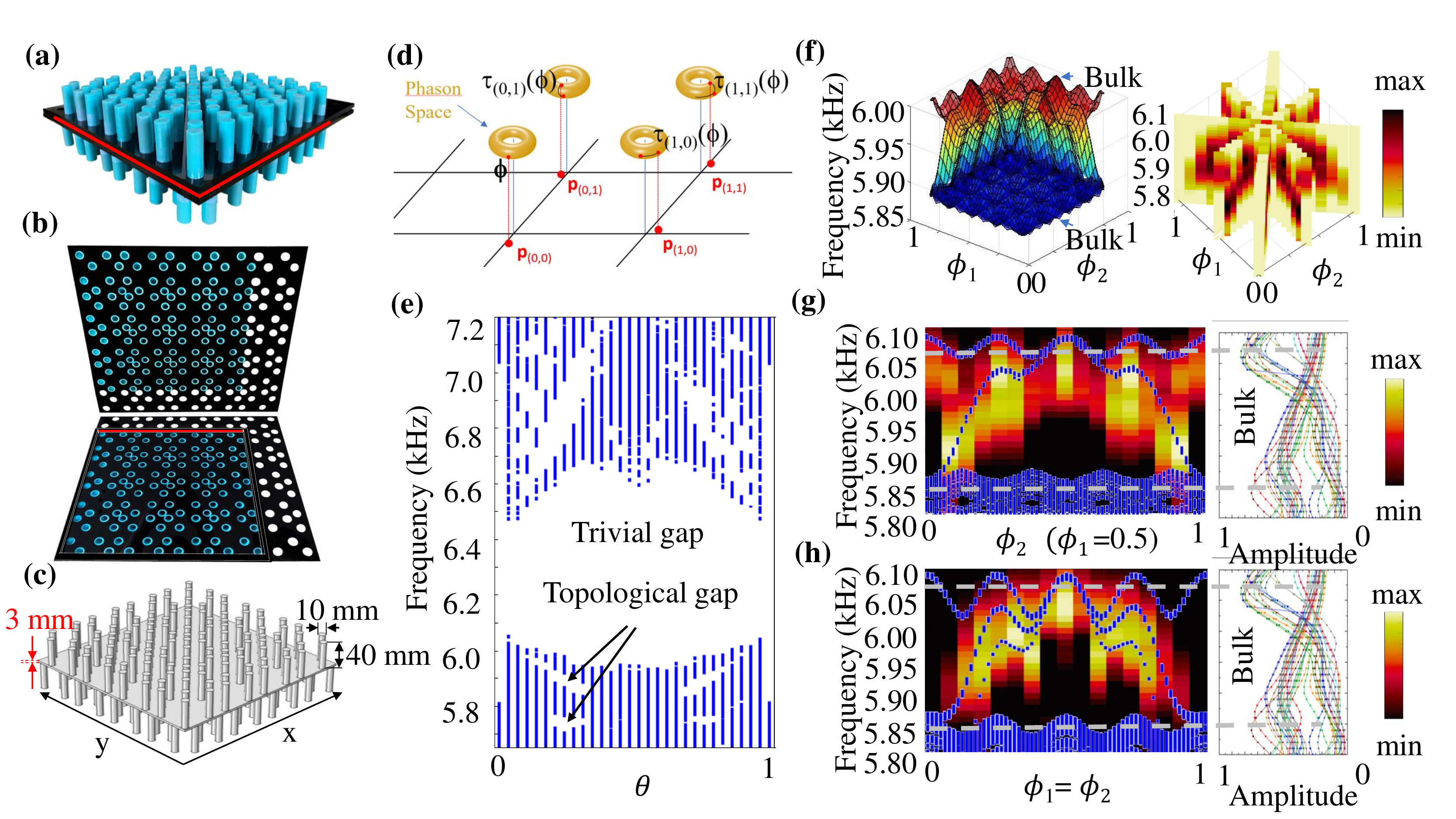}
\caption{\small {\bf Fig 1 4D Topological Quantum Hall Effect.} {\bf a} Photograph of a fully assembled acoustic 2D sinusoidal pattern consisting of top/bottom cylindrical resonators. The middle red bar indicates the presence of an inner chamber, which connects the top and bottom resonators and is referred to as the spacer. {\bf b} Photograph of the inner structure, with the spacer now fully visible. {\bf c} The wave propagation domain, together with relevant parameters. {\bf d} Illustration of the algorithm which supplies the position of the resonators. {\bf e} COMSOL simulated bulk resonant spectrum against $\theta$, together with labels for the topological and non-topological gaps. {\bf f} Left: COMSOL simulated resonant spectrum for hard wall termination, shown against the phason parameters $\phi_1$ and $\phi_2$. The Weyl singularity is the spectral surface connecting the the indicated bulk bands. Right: Experimental measurement of the density of states, with the phason space sampled in several directions. {\bf g, h} Comparison between the experimentally measured density of states and the simulated spectrum (blue dots) for the traces $\phi_1=0.5$ and $\phi_1 = \phi_2$, respectively. The theoretical spectra have been stretched by a small factor to overlay the experimental data. The bright dispersive modes in the left side are part of the spectral dome. The right side displays the raw density of states data for the bulk, from where we read the edges of the bulk gap.} 
\label{Fig1}
\end{figure*}

These three works\cite{LohseNature2018,ZilberbergNature2018,WangNaatComm2020}  are the only experimental emulations of the 4D QHE to date, and many aspects related to the effect remained un-confirmed. Certainly, an experimental setup where both the spatial and frequency domains can be simultaneously resolved is missing. Even with the aperiodic principles at hand, the designs remained challenging, perhaps because it is falsely assumed that the models need to simulate the Hofstadter Hamiltonians as close as possible. However, as pointed out by Apigo {\it et al},\cite{ApigoPRM2018} no fine tuning is actually necessary to open topological gaps. The latter only require strong aperiodicity and strong couplings between the resonators, as well as a reliable strategy for the phason engineering that produces the desired topological phases.\cite{ProdanJGP2019} Guided by these principles, we demonstrate here a robust design of a quasi-periodic 2D acoustic crystal that hosts the 4D quantum Hall physics. Reconfigurability and other advantages of the experimental setup enables us to map, the Weyl singularity predicted by the bulk-boundary correspondence.\cite{ProdanSpringer2016} Furthermore, we demonstrate ways to control and steer the boundary modes using the phason, that are specific only to 4D QHE.

All the above rely on the concept of phason engineering. Using this strategy, the Hall physics from an arbitrarily high dimension can be access from a physical space of lower dimension $d=1,2,3$. The high throughput of large classes of topological models produced by phason engineering will be essential for our understanding of higher-dimensional bulk-boundary correspondence, of its manifestation in meta-materials and of its possible applications. Phason engineering relies on a specialized algorithm to position the resonators relative to each other to produce phasons that live on arbitrary $d'$-dimensional tori. Regardless of the particular couplings of the resonators, the dynamical matrices that determine the collective resonant spectrum of the crystal are Galilean invariant. We show that any such Galilean invariant dynamical matrix is just linear combinations and products of elementary operators that satisfy the commutation relations of the magnetic translations in $(d+d')$-dimensions. As such, the spectral gaps of the crystals carry higher Chern numbers and they display a bulk-boundary correspondence specific to IQHE in higher dimensions.\cite{ProdanSpringer2016} Furthermore, to navigate the complex topology of the states, we devise a K-theoretic visual method to map the large number of topological invariants associated with the bulk gaps, based on the gap labeling technique.\cite{Bellissard1986,Bellissard1995}.

\section{Experimental set-up and results} 

Our experimental set-up and main results are summarized in Fig.~\ref{Fig1}. Photographs of our acoustic crystal can be seen in Fig.~\ref{Fig1}(a,b). It consists of identical cylindrical resonators with a geometry specified in Fig.~\ref{Fig1}(c). The geometry was chosen to accommodate the microphone and the speaker used for the measurements (see Methods), as well as to ensure a good separation of the discrete resonant modes, in the frequency domain we are interested in. The latter ensures that only basic mode-to-mode couplings occur. Figure~\ref{Fig1}(c) also shows the domain of the acoustic wave propagation and, as one can see, the resonators are connected through a thin domain, which we call the spacer. It is highlighted in red in Fig.~\ref{Fig1}(a). By filling this spacer with solid material, we can confine the wave propagation and create a re-configurable boundary, highlighted in red in Fig.~\ref{Fig1}(b), to control the phason as explained below. It is important to acknowledge that the use of a spacer as a solution for resonator coupling, and not other complicated means such as bridges, is one of the keys to our results. The spacer, certainly, does not allow any fine tuning but it does enable strong coupling, hence it is of crucial importance that accessing the 4D Hall physics does not rely on fine tuning.\cite{ApigoPRM2018} Let us specify that, in order to tightly pack the resonators and increase the strength of the coupling, the nearest neighboring resonators have been placed on opposite sides of the spacer.

\begin{figure}[t!]
\includegraphics[width=\linewidth]{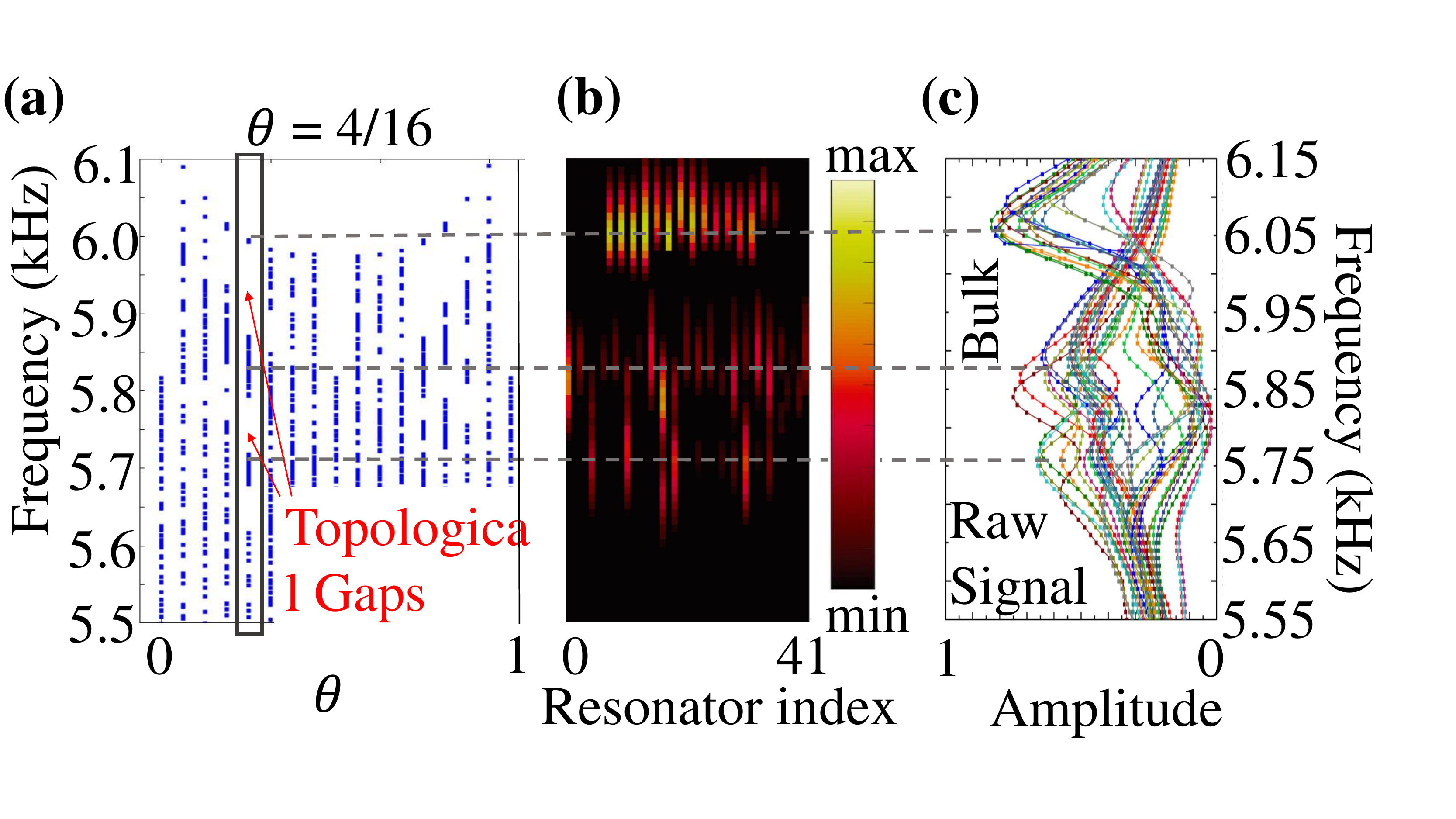}
\caption{\small {\bf Fig 2 Bulk measurements}. {\bf a} COMSOL simulated resonant spectrum for the experimental set-up from Fig.~\ref{Fig1}, with arrows indicating the topological gaps. The vertical box identifies $\theta = 0.25$, used in experiments. {\bf b} Measured local density of states, assembled from microphone readings on 42 bulk resonators. {\bf c} Collapse on the frequency axis of the intensity plot reported in panel {\bf b}. Two spectral gaps can be clearly identified and seen to be well aligned with the theoretical predictions.}
\label{Fig2}
\end{figure}

The resonators are labeled by $\bm n =(n_1,n_2) \in \ZM^2$ and the $(x,y)$-coordinates of their centers $\bm p_{\bm n}$ are such that
\begin{equation}\label{Eq:Pattern1}
\begin{aligned}
&\bm p_{\bm n+(1,0)} = \bm p_{\bm n} + D \, \Big ( 1 +  \epsilon \sin\big (2 \pi(\phi_1+n_1 \theta)\big), 0 \Big ),
\\
& \bm p_{\bm n+(0,1)} = \bm p_{\bm n} + D \, \Big ( 0, 1 + \epsilon \sin \big (2 \pi(\phi_2 + n_2 \theta) \big ) \Big ),
\end{aligned}
\end{equation} 
with $D=17.0$~mm and $\epsilon=0.4$. This pattern can be thought as a strongly perturbed ideal lattice with the perturbation produced dynamically, via the specialized algorithm illustrated in Fig.~\ref{Fig1}(d) and explained below. Note that the pattern depends on the phason $\bm \phi = (\phi_1,\phi_2)$, which lives on a 2-torus. We place the boundaries along the horizontal axis $x = - D/2$ and vertical axis $y= - D/2$, hence in between the rows and columns of the ideal lattice. Now, if we keep the pattern in place and we move the boundaries at $x=(m_1-\frac{1}{2})D$ and $y=(m_2-\frac{1}{2})D$, then this move has the same effect as changing the phason as $\bm \phi \mapsto (\phi_1+m_1 \theta,\phi_2+m_2 \theta)$. As such, by having an adjustable boundary, we can sample the phason space with just one acoustic crystal. In the actual experiments, we used four different acoustic crystals together with the mobile boundary technique to sample 16$\times$16 points of the phason space. Let as acknowledge that Eq.~\eqref{Eq:Pattern1} is just one example of a large class of patterns generated with the specialized algorithm illustrated in Fig.~\ref{Fig1}(d), which give access to the 4D quantum Hall physics (see Phason engineering). 

The resonant spectrum of the acoustic crystal, as computed with the finite-element based software COMSOL \cite{COMSOL}, is reported in Fig.~\ref{Fig1}(e) as function of $\theta$. Since the simulations were for a finite crystal, some of the bulk gaps are contaminated by boundary spectrum. Additional model calculations with periodic boundary conditions and for larger crystals are reported in \emph{Supplementary Information}. At $\theta=0$, the crystal is periodic and, as expected, the spectrum contains bands that evolve from the discrete modes of the individual resonators. These bands don't share any dynamical features, hence the spectral gaps separating them are all trivial [see the label in Fig.~\ref{Fig1}(e)]. As the parameter $\theta$ is turned on, the bands of spectrum become fragmented and a large number of spectral gaps develop. Qualitatively, the spectra resembles the Hofstadter butterfly \cite{HoftadterPRB1976} and, as we shall see below, the spectral gaps carry 2$^{\rm nd}$ and 1$^{\rm st}$ Chern numbers. 

\begin{figure}[t!]
\includegraphics[width=\linewidth]{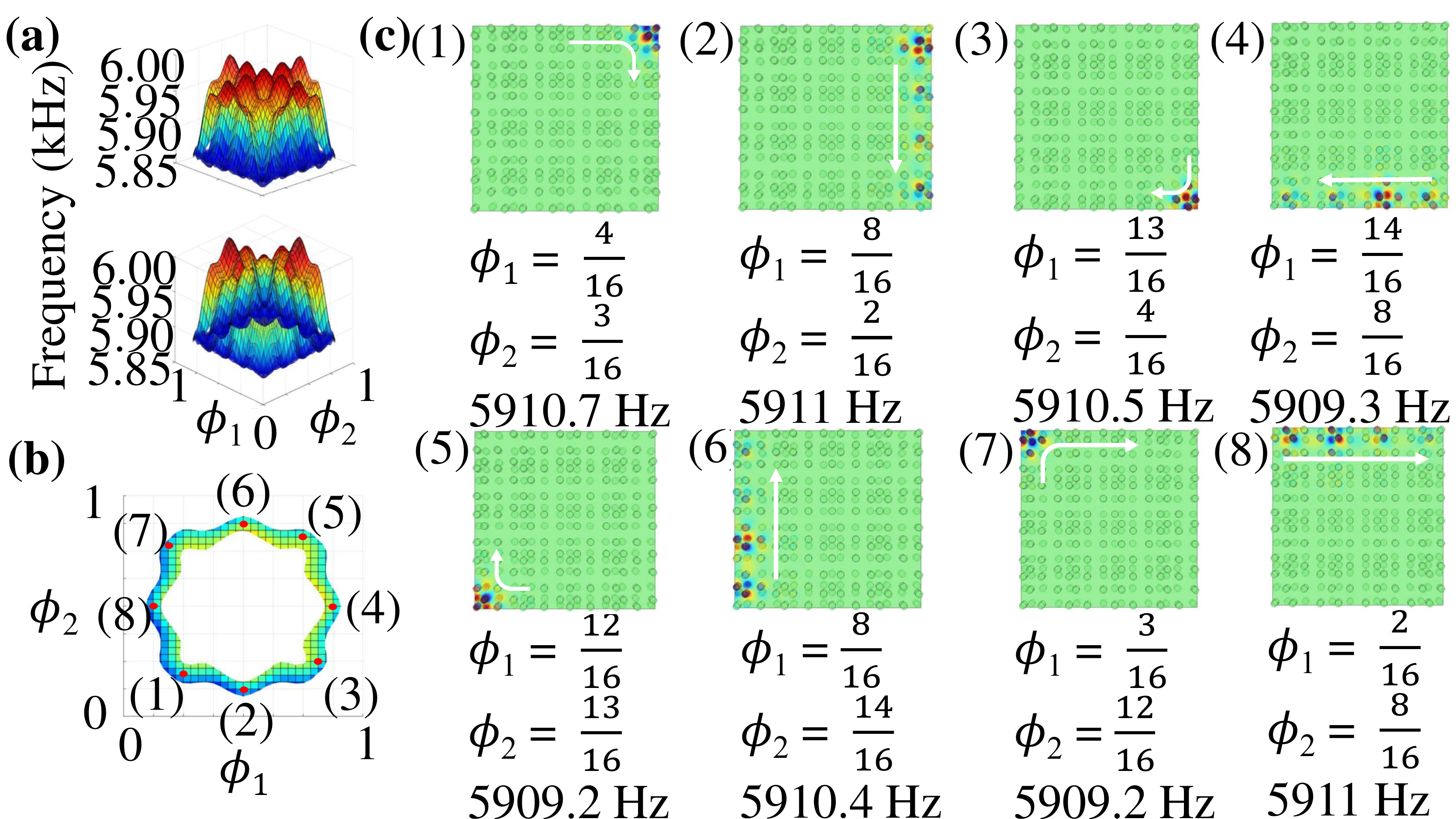}
\caption{\small  {\bf Fig 3 Weyl singularity and mode steering.} {\bf a} COMSOL simulated spectrum as function of the phason $\bm \phi =(\phi_1,\phi_2)$, revealing the Weyl singularity. {\bf b} Cross-section of the Weyl singularity at 5910~Hz. {\bf c} Acoustic pressure field distribution for the eight phasons marked in panel {\bf b}, revealing circular mode steering around the crystal's boundary.}
\label{Fig3}
\end{figure}

Experimentally, we were able to reproduce with high fidelity the predicted spectra from Fig.~\ref{Fig1}(e), as demonstrated in Fig.~\ref{Fig2}. Specifically, in Fig.~\ref{Fig2}(b), we report the measured local density of states of the crystal, resolved by frequency and resonator index (see Methods). The data is collapsed on the frequency axis in Fig.~\ref{Fig2}(c) and two clear spectral gaps are identified, which are well aligned with the ones in the COMSOL simulated spectrum, shown again in Fig.~\ref{Fig2}(a). The resonant spectrum in Fig.~\ref{Fig2}(a) does not depend on $\phi_1$ and $\phi_2$.  In the following, we fix $\theta$ at the value identified in Fig.~\ref{Fig2}(a) and work with the first bulk gap, counted from the top, which as we shall see, carries  a 2$^{\rm nd}$ Chern number ${\rm Ch}_2=-1$.

In Fig.~\ref{Fig1}(f), we report on the right the resonant spectrum of a finite crystal as function of $\bm \phi$, as computed with COMSOL in a finite frequency domain that covers the bulk gap identified above. The dominant feature connecting the indicated bulk bands of spectrum is a spectral dome that is hollow inside. A point inside this dome has no escape path since it is completely surrounded by spectrum. Furthermore, using model calculations, we observed that this dome does not disappear or open up under continuous deformations of the crystal. These indicate that the dome has a built in topological protection, which we will associate with the $2^{\rm nd}$ Chern bulk number and with the Weyl physics expected at the surface of a 4D IQHE system. Let us state that we were able to experimentally reproduce with high fidelity this spectral dome. Indeed, the right side of Fig.~\ref{Fig1}(f) reports the experimentally measured density of states (see Methods), with the phason space sampled in several directions and, as one can see, the outcome reproduces the spectral dome. In Fig.~\ref{Fig1}(g,h) we report two sections of these measurements showing quantitative agreement with the COMSOL simulations. The Weyl singularity is further analyzed in Fig.~\ref{Fig3}. In particular, it is shown there that the modes associated to that part of the spectrum are localized on the boundary. Experimental measurements of the spatial profiles of these modes, confirming the boundary localization, are reported in \emph{Supplementary Information}. 

The exactly solved 4D IQHE model in \cite[Sec.~2.2.4,]{ProdanSpringer2016}was isotropic in all four dimension and, in the presence of a flat boundary, the dispersion was found to display a Weyl singularity $E(\vec k_\|) = \pm \|\vec k_\|-\vec k_0\|$, where $\vec k_\|$ are the three quasi-momenta parallel to the boundary and $\vec k_0$ is the coordinate of the Weyl singularity. Our acoustic crystal is highly isotropic and the Weyl singularity for a flat boundary is collapsed, in the sense that the boundary spectrum displays dispersion only with respect to $\phi_1$ if the boundary is cut perpendicular to the first spatial direction. Nevertheless, the Weyl physics of the boundary is still encoded (see \cite[Example 5.3.3)]{ProdanSpringer2016}in the boundary topological invariant supplied by the 3-dimensional winding number of the gap unitary operator $U_G(\bm \phi) = e^{\imath 2 \pi g(D_{\bm \phi})}$, where $D_\phi$ is the dynamical matrix of the crystal with a boundary and $g$ is any continuous real valued function taking 0/1 value above/below the bulk gap. By construction, the spectral decomposition of $U_G-I$ involves only the boundary modes that are spatially localized at the edges of the sample. Now, our sample has four edges instead of just a flat one, hence, for the sake of the argument, it is more appropriate to consider a disk-shaped sample of very large radius $R$. Then the winding number is computed using the variables $\phi_1$, $\phi_2$ and the momentum $k_\|$ parallel to boundary, with the latter treated with the real-space methods.\cite{ProdanSpringer2017} Specifically, the derivation $\partial_{k_\|}(\cdot)$ is replaced by the commutator $\imath [\cdot,R \hat \phi_3]$ and $\int d k_\|(\cdot)$ is replaced by $\frac{1}{2\pi R}{\rm Tr}(\cdot)$, where $\hat \phi_3$ is the operator corresponding to the polar angle in the plane of the sample. The radius $R$ cancels out and the 3D winding number takes the form
\begin{equation}\label{Eq:W3}
W_3(U_G) = \Lambda_3 \sum_{\sigma \in \Ss_3} (-1)^\sigma \int {\rm d}^3 \bm \phi \, \prod_j U_G^{-1} \partial_{\sigma_j}U_G,
\end{equation}
where $\Lambda_3$ is the standard normalization constant and $\Ss_3$ is the group of permutation of three objects.
 The bulk-boundary correspondence (see \onlinecite[Sec. 5.5)]{ProdanSpringer2016}assures that this boundary invariant equals the $2^{\rm nd}$ Chern number of the bulk gap projection \cite{Footnote1} and, as such, a Weyl singularity is expected if the modes are resolved by $\phi_j$'s. The dome observed in Fig.~\ref{Fig1}(f) carries the boundary invariant~\eqref{Eq:W3} and, for this reason, we proclaim that this spectral feature is the manifestation of the Weyl physics expected at the boundary of a 4D IQHE system. In particular, the spectral dome cannot open in any spatial direction. Such feature is expected for more general boundaries and we have verified this statement for a sample shaped like an octagon (see Fig.~\ref{Fig:SpectralDome}). 

In \cite[]{ZilberbergNature2018} light was injected in a corner of the ensemble of modulated wave guides and light was observed coming out at the opposite corner. It was inferred that their observation is equivalent to adiabatic pumping along the cycle mapped in Fig.~\ref{Fig1}(h) (see also \cite{ZPRR2020}). We were indeed able to reproduce this interesting  corner-to-corner pumping effect (see \emph{Supplementary Material}), but let us point out that this type of pumping occurs through the bulk states. In an actual pumping experiment, this will inherently lead to leakage into the bulk modes. The Weyl singularity, however, gives access to additional pumping cycles that avoid the bulk spectrum. Indeed, one of the special features of a strong topological invariant, such as the 2$^{\rm nd}$ Chern number, is that boundary modes occur regardless of the orientation of the boundary. This feature, together with the full control over the phason, enable a corner-to-corner mode steering that does not proceed through the bulk states but rather goes around the Weyl singularity, as well as around the edges of the sample. The effect is illustrated in Fig.~\ref{Fig3}, were a section of the Weyl singularity was sampled at eight points in Fig.~\ref{Fig3}(b) and the spatial profiles of the corresponding eigenmodes were mapped in Fig.~\ref{Fig3}(c). As one can see, the mode is steered around the boundary of the crystal and completes a full cycle as the phason is cycled over the section of the Weyl singularity.

\section{Methods}

\subsection{Fabrication} Our fabrication process is modular and the acoustic crystals are assembled from parts that are independently manufactured with different automated process. This approach enables a high throughput of acoustic crystals, which can be dis-assembled and stored after use. 

One leg of the process is the manufacturing of the supporting bases, which are x-mm thick acrylic plates with through holes, laser-cut with the Boss Laser-1630 Laser Engraver. A specialized piece of computer code communicates the pattern~\eqref{Eq:GenAlg} and the radius of the holes to the machine, hence different phason designs can be efficiently implemented. The nominal tolerance of the laser-cutter is 250 um. 

The resonators were manufactured using an Anycubic Photon 3D printer, which uses UV resin and has 47~um XY-resolution and 10~um Z-resolution. The thickness of their walls is 2~mm, to ensure a good quality factor and to justify rigid boundaries in our numerical simulations. The inner dimensions of the resonators are supplied in Fig.~\ref{Fig1}. Identical resonators were printed in large quantities and were made ready for the assembling.

The resonators were pushed through the holes of the base plates until flushed with the opposite side of the acrylic plates. After the top and bottom parts were fully assembled, they were pressed against the spacer, which is a 3~mm thick acrylic plate with a large opening cut out to accommodate a total of $16 \times 16$ resonators. Let us specify that top and bottom parts accommodate a total of 23$\times$23 resonators, such that spacer can be moved around and generate crystals with different phasons, as explained in the main text.

\subsection{Experimental protocols} The protocol for the acoustic bulk and edge measurements reported in Fig.~\ref{Fig1} h and Fig.~\ref{Fig2} was as follows: Sinusoidal signals of duration 1~s and amplitude of 0.5~V were produced with a Rigol DG 1022 function generator and applied on a speaker placed in a porthole opened in a resonator of the bottom row. A dbx RTA-M Reference Microphone with a Phantom Power was inserted in a porthole opened in a resonator of the top row and acquired the acoustic signals. To account for the frequency-dependent response of the components, several separate measurements were performed with the structure removed but speaker and microphone kept at the same positions. All our microphone readings are normalized by these reference measurements. The signals were read by a custom LabVIEW code via National Instruments USB-6122 data acquisition box and the data was stored on a computer for graphic renderings.  

The local density of states reported in Fig.~\ref{Fig2}(b) was acquired with the above protocol, which was repeated for 42 resonators located away from the boundary. For each resonator, the frequency was swept over the shown range of frequency. According to the acoustic pressure field distribution of the upper and lower body modes of the topological gap in the COMSOL simulation, the microphones was placed on 42 different resonators with measurable acoustic pressure. The speaker was placed on the nearest resonator immediately below the microphone and, as a result, it was moved between the bottom resonators.

The density of states reported in Fig.~\ref{Fig1} (g) and (h) was obtain by integrating the local density of states acquired from resonators close to the boundary. Same instrumentation was used but the microphone was inserted in the resonator along the boundary according to the acoustic pressure distribution of the edge modes in the COMSOL simulation, and the speaker was placed on the nearest resonator immediately above the microphone. In Fig.~\ref{Fig1} (g), from 0 to $0.5$, the acoustic pressure field distribution of the edge mode is concentrated on the right and upper boundary, and from $0.5$ to $1$, the acoustic pressure field distribution of the edge mode is concentrated on the left and lower boundary. In Fig.~\ref{Fig1} (h), from 0 to $0.5$, the acoustic pressure field distribution of the edge mode is concentrated on the right and upper boundary and upper right corner, and from $0.5$ to $1$, the acoustic pressure field distribution of the edge mode is concentrated on the left and lower boundary and lower left corner. The measurements were repeated with the change of phason in steps of $1/16$. For each measurement, the frequency was scanned from 5800~Hz to 6100~Hz in 10~Hz steps. The full map of the density of states reported in Fig.~\ref{Fig1} (f) was obtained by the assembling the data from Fig.~\ref{Fig1} (g) and (h) and symmetry considerations to fill in the data for the two additional directions shown in Fig.~\ref{Fig1} (f).

\section{Simulation}
The simulations reported in Figs.~\ref{Fig1}, \ref{Fig2} and \ref{Fig3} were performed with the COMSOL Multiphysics pressure acoustic module. The domain shown in Fig.~\ref{Fig1}{c} was filled with air with a mass density 1.3~kg/m$^{3}$ and the sound's speed was fixed at 343 m/s, appropriate for room temperature. Because of the huge acoustic impedance mismatch compared with air, the 3D printing UV resin material was considered as hard boundary.

The spectra reported in Fig.~\ref{Fig:SpectralB} were computed with the stated model dynamical matrices, which were coded in Fortran and exactly diagonalized using the standard LAPACK library. The simulations assumed 101$\times$101 resonators and $\theta$ was sampled as $n/101$ for panels (a,c) and as $\sqrt{2}n/101$ for panel (b), $n=\overline{1,101}$. This particular sampling enabled us to impose boundary conditions.

\begin{figure}[t!]
\includegraphics[width=\linewidth]{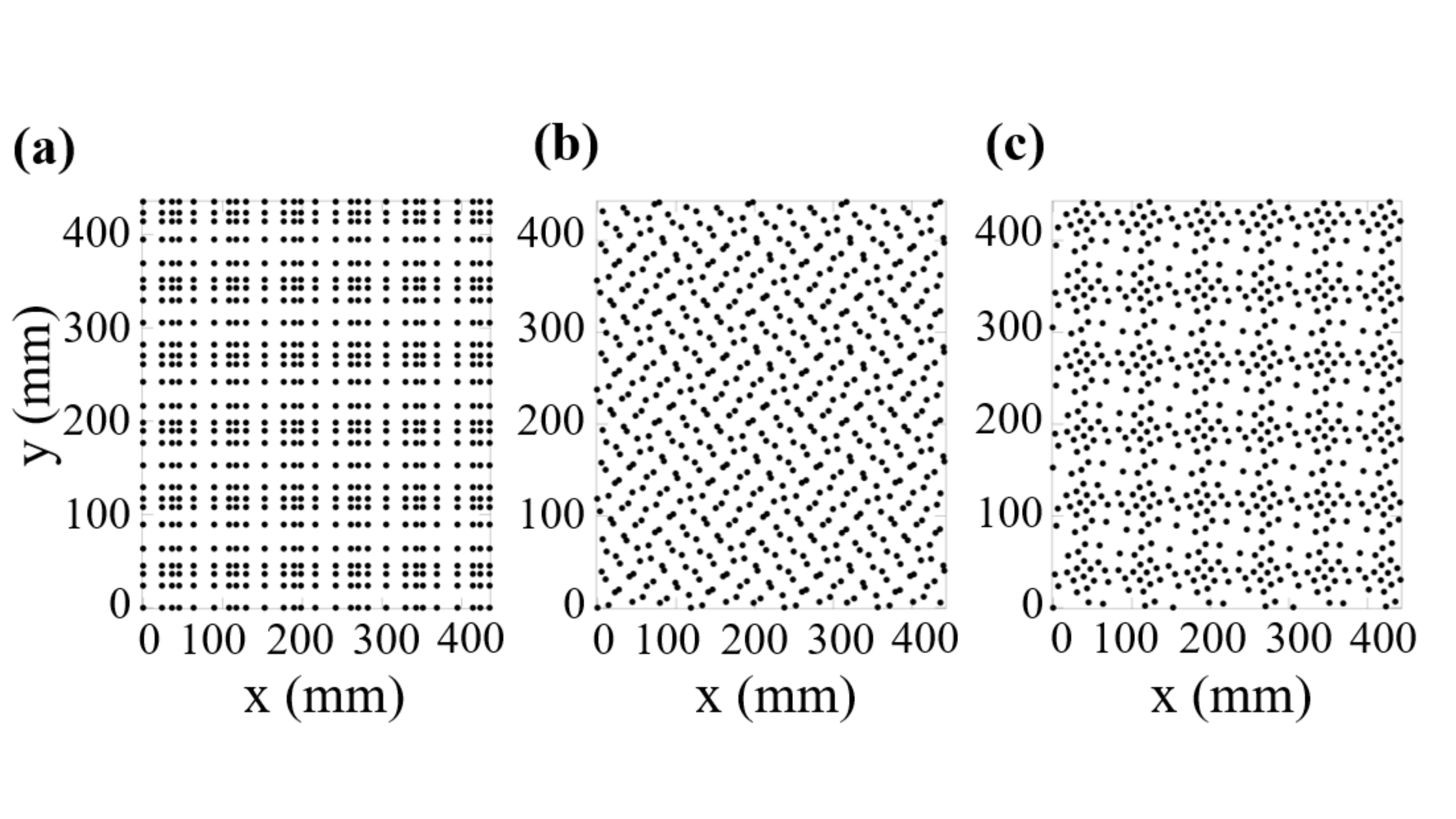}
\caption{\small  {\bf Fig 4 Dynamically generated patterns.} {\bf a} Same as the pattern of resonators in Fig.~1 but with the simplified $F(\bm \phi) = \epsilon D\big (\sin(2 \pi \phi_1),\sin(2 \pi \phi_2)\big )$. {\bf b} Same as {\bf a} but with the lattice $\Ll'$ rotated by $45^\circ$ relative to $\Ll$. {\bf c} Same as $\bf a$ but with $F$ replaced by $G \circ F$ with $G(x,y) = (x+y,y-x)$. All three patterns were generated with $\theta = 1/2\sqrt{2}$.}
\label{Fig:Patterns}
\end{figure}

\section*{Phason engineering} 
We show here that, by using a specialized algorithm to position the resonators relative to each other, we can engineer phasons that live on arbitrary $d'$-dimensional tori. The algorithm starts from a Bravais lattice $\Ll$, generated by acting on the origin $\bm p_0$ of the physical space $\RM^d$ ($d=1,2,3$) with an abelian group of discrete translations 
\begin{equation}
\mathfrak t_{\bm n}\bm x = \bm x+\sum n_i \bm a_i, \ \bm n = (n_1,\ldots, n_d) \in \ZM^d.
\end{equation} 
The second ingredient is a virtual space $\RM^{d'}$ with $d' \geq d$ and with $\RM^d$ canonically embedded in $\RM^{d'}$. Lastly, $\Ll'$ is an independent Bravais lattice inside the virtual space $\RM^{d'}$, generated by $\bm a'_j$, $j=\overline{1,d'}$. With these ingredients in hand, we are going to produce a topologically ergodic dynamical system, as the one depicted in Fig.~\ref{Fig1}(d). For this, we form the $d'$-torus $\TM^{d'} = \RM^{d'}/\Ll'$ and let the abelian group $\ZM^d$ act on it via shifts
\begin{equation}
\tau_{\bm n}(\bm \phi) = \Big (\bm \phi + \sum_{i=1}^d n_i \bm a_i \Big){\rm mod}\, \Ll', \quad \bm \phi \in \TM^{d'},
\end{equation}
exactly as in Fig.~\ref{Fig1}(d). Then, if $F: 
\TM^{d'} \rightarrow \RM^d$ is any continuous map, such as the projection in Fig.~\ref{Fig1}(d), we generate the quasi-periodic pattern $\{p_{\bm n}(\bm \phi)\}_{\bm n \in \ZM^d}$ in $\RM^d$ using the algorithm
\begin{equation}\label{Eq:GenAlg}
\bm p_{\bm n}(\bm \phi) = \mathfrak t_{\bm n}\Big ( \bm p_0  + F\big (\tau_{\bm n}(\bm \phi)\big)\Big ), \quad \bm \phi \in \TM^{d'}, \ \bm n \in \ZM^d,
\end{equation}
which is graphically illustrated in Fig.~\ref{Fig1}(d). Note, for example, that point $\bm p_{(1,1)}$ in Fig.~\ref{Fig1}(d) can be reached from the origin using many other different paths, such as $(0,0) \mapsto (0,1) \mapsto (1,1)$. For the algorithm to be well defined, all these many different paths must produced the same point $\bm p_{(1,1)}$. This is indeed ensured by the fact that $\tau$ is a group action, that is, $\tau_{\bm n}\tau_{\bm m} = \tau_{\bm n + \bm m}$. This property is highly non-trivial as, for example, if one chooses to replace the torus with a sphere, then one will find that there are no natural actions of $\ZM^d$ on such space, besides the trivial cyclic ones.

\begin{figure}[t!]
\includegraphics[width=\linewidth]{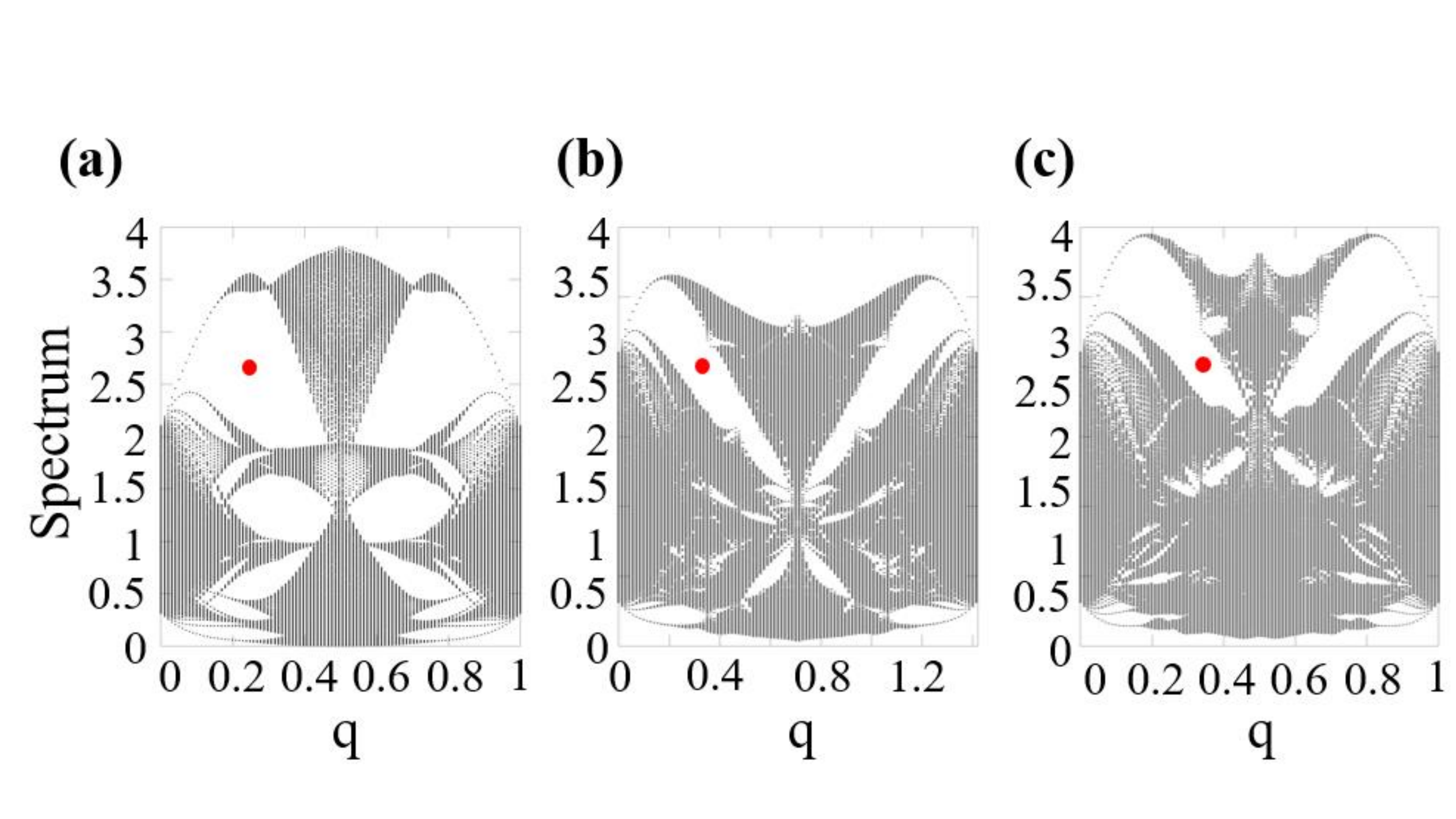}
\caption{\small  {\bf Fig 5 Spectral butterflies.} Panels ({\bf a},{\bf b},{\bf c}) correspond to the patterns in  Figs.\ref{Fig:Patterns}({\bf a},{\bf b},{\bf c}), respectively. The spectra were computed with the model dynamical matrix $D_\Pp = \sum_{\bm x,\bm y \in \Pp} e^{-1.5|\bm x -\bm y|^2 }\, |\bm x\rangle \langle \bm y |$, with the distance measured in units of $D$. The marked spectral gaps are related to bulk spectral gaps mapped in the experiments.}
\label{Fig:SpectralB}
\end{figure}

The pattern we just designed in the physical space, $\Pp_{\bm \phi} = \{\bm p_{\bm n}(\bm \phi)\}_{\bm n \in \ZM^d}$, depends on the phason $\bm \phi$, which marks the point where the algorithm is started. Furthermore,  rigid shifts of the pattern result in re-adjustments of the phason, $
\mathfrak t_{\bm a}^{-1}\Pp_{\bm \phi} = \Pp_{\tau_{\bm a}\bm \phi}$. Since rigid shifts do not change the resonant spectrum of a crystal, we arrive at the crucial conclusion that the spectrum of the dynamical matrix $D_{\bm \phi}$ is {\it independent} of the phason, provided the orbit $\tau_{\bm a} \bm \phi$ densely fills the $d'$-torus. As it is well known, this is indeed the case if the lattices $\Ll$ and $\Ll'$ are incommensurate. To conclude, we just showed how to engineer a pattern with a phason that lives on a $d'$-torus. The latter can be used as an adiabatic parameter that keeps {\it every single bulk spectral gap} open, regardless of how the phason is cycled over $\TM^{d'}$. This is paramount for the existence and protection of the dispersive boundary modes.

For the acoustic crystal shown in Figs.~\ref{Fig1}, the center of the resonators were positioned as in \eqref{Eq:GenAlg}, with $d=d'=2$, $\bm p_0$ at the lower-left corner and
\begin{equation}
F(\bm \phi)= - \frac{\epsilon D}{2\sin(\pi \theta)} \, \big (\cos(2 \pi \phi_1+\pi\theta),\cos(2 \pi \phi_2+\pi \theta)\big ),
\end{equation} 
where $\bm \phi = \phi_1 \bm a'_1 + \phi_2 \bm a'_2$, $\phi_i \in [0,1]$. Also $\Ll$ and $\Ll'$ are square lattices related as $\Ll = \theta \Ll'$. The 4D quantum Hall physics, however, can be accessed with any other function $F$ or lattice $\Ll'$. Fig.~\ref{Fig:Patterns} supplies such examples and, as one can see, the texture of the pattern can change drastically when such adjustments are considered. As we shall see below, as long as a bulk gap remains opened, the topological boundary spectrum shown in Fig.~\ref{Fig1} cannot be removed under smooth deformations of either $F$ or $\Ll'$. 

\begin{figure}[t!]
\includegraphics[width=\linewidth]{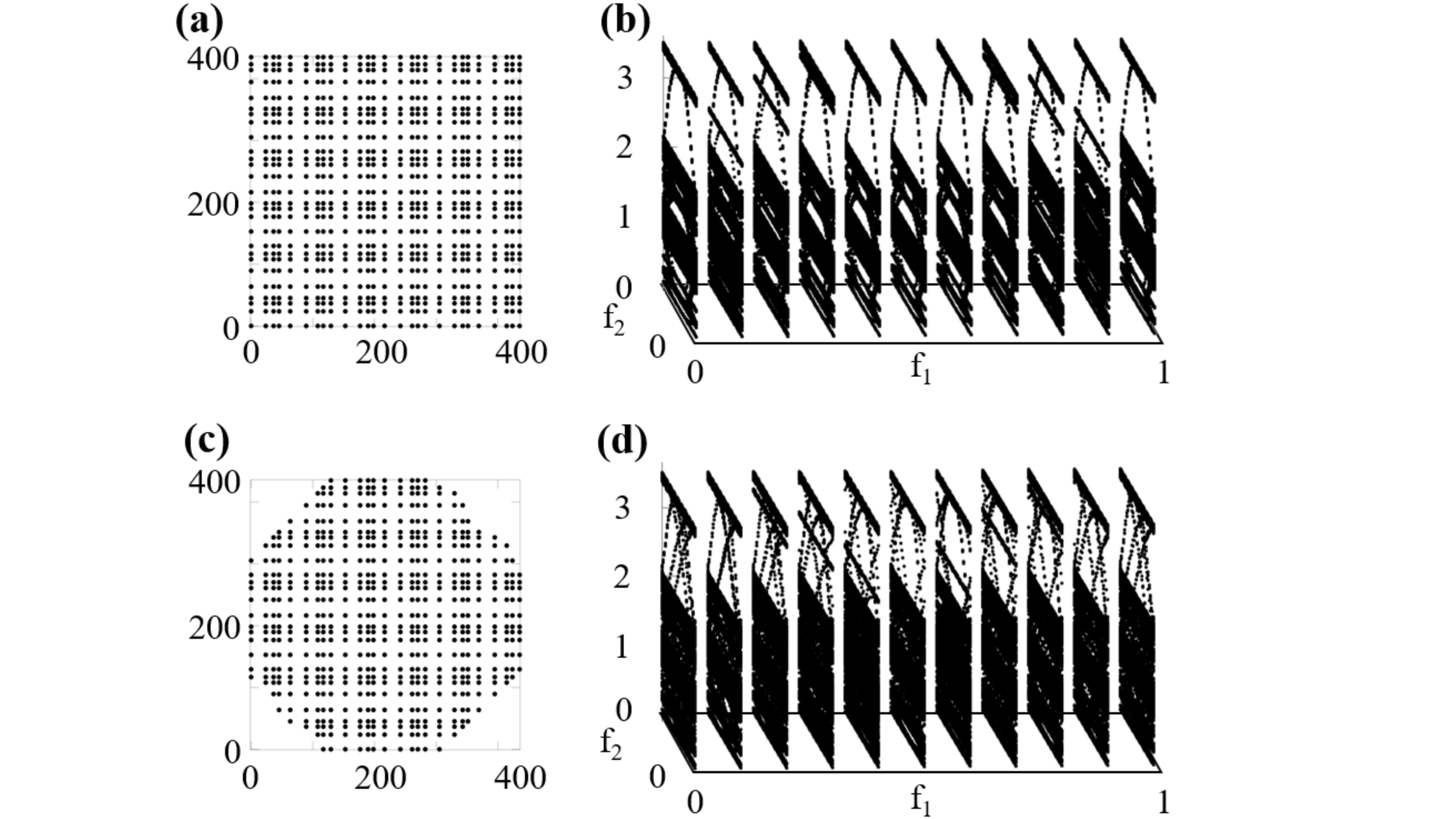}
\caption{\small  {\bf Fig 6 Stability of the spectral dome.} {\bf a} A crystal cut as a square and {\bf b} its resonant energy spectrum. {\bf c} A crystal cut as an octagon and {\bf d} its resonant energy spectrum. In both cases, the spectra have been resolved by $\phi_1$ and $\phi_2$ and a spectral dome can be clearly identified. The spectra were computed with the model {\bf a} from Fig.~\ref{Fig:SpectralB}.}
\label{Fig:SpectralDome}
\end{figure}

\section{QHE via phason engineering} 
We now demonstrate how the quantum Hall physics emerges in these systems. For mode-to-mode coupling, we can focus on one spectral band at a time, say the $k$-th one. Then the dynamical matrix corresponding to a generic pattern $\Pp$ of resonators takes the form
\begin{equation}\label{Eq:GenDyn}
D_{\Pp} = \sum_{\bm m,\bm n \in \ZM^d} w_{\bm m,\bm n}(\Pp) |\bm m\rangle \langle \bm n|,
\end{equation}
where $|\bm n\rangle$ encodes the $k$-th discrete resonant mode of the resonator placed at position $\bm p_{\bm n}$. As alluded by the notation, the overlap parameters $w_{\bm m,\bm n}(\Pp)$ must depend on the pattern in a continuous fashion and they must obey the constraints
\begin{equation}\label{Eq:Galilean}
w_{\bm m,\bm n}(\mathfrak t_{\bm a}\Pp) =  w_{\bm m + \bm a,\bm n + \bm a}(\Pp), \quad \bm a \in \ZM^2.
\end{equation}
It is important to acknowledge that there is no more general expression than \eqref{Eq:GenDyn}, because $\Pp$ encodes the entire geometric data of the crystal. Also, Eq.~\eqref{Eq:Galilean} follows entirely from the Galilean invariance. For our specific patterns, we can pass from $\Pp$ to $\bm \phi$ and use Eq.~\ref{Eq:Galilean} to reduce
\begin{equation}
D_{\bm \phi} =\sum_{\bm q} S_{\bm q} \Big ( \sum_{\bm n} w_{\bm q,\bm 0}\big(\tau_{\bm n}(\bm \phi)\big) \, |\bm n \rangle \langle \bm n| \Big ),
\end{equation}
where $S_{\bm q}|\bm n\rangle = |\bm n +\bm q\rangle$ is the shift operator. This shows that any Galilean invariant dynamical matrix over the pattern $\Pp_{\bm \phi}$ is a combination of shift operators and diagonal operators of the form
\begin{equation}
T_f = \sum_{\bm n} f\big(\tau_{\bm n}(\bm \phi)\big) \, |\bm n \rangle \langle \bm n|,
\end{equation}
for some continuous function $f$ over $d'$-torus. Furthermore, the following commutation relations are obvious
\begin{equation}
 T_{f} \, S_{\bm q} = S_{\bm q} \, T_{f \circ \tau_{\bm q}}, \quad T_f \, T_g = T_g \, T_f = T_{f\cdot g}.
\end{equation}

\begin{figure*}[t!]
\center
\includegraphics[width=0.8\linewidth]{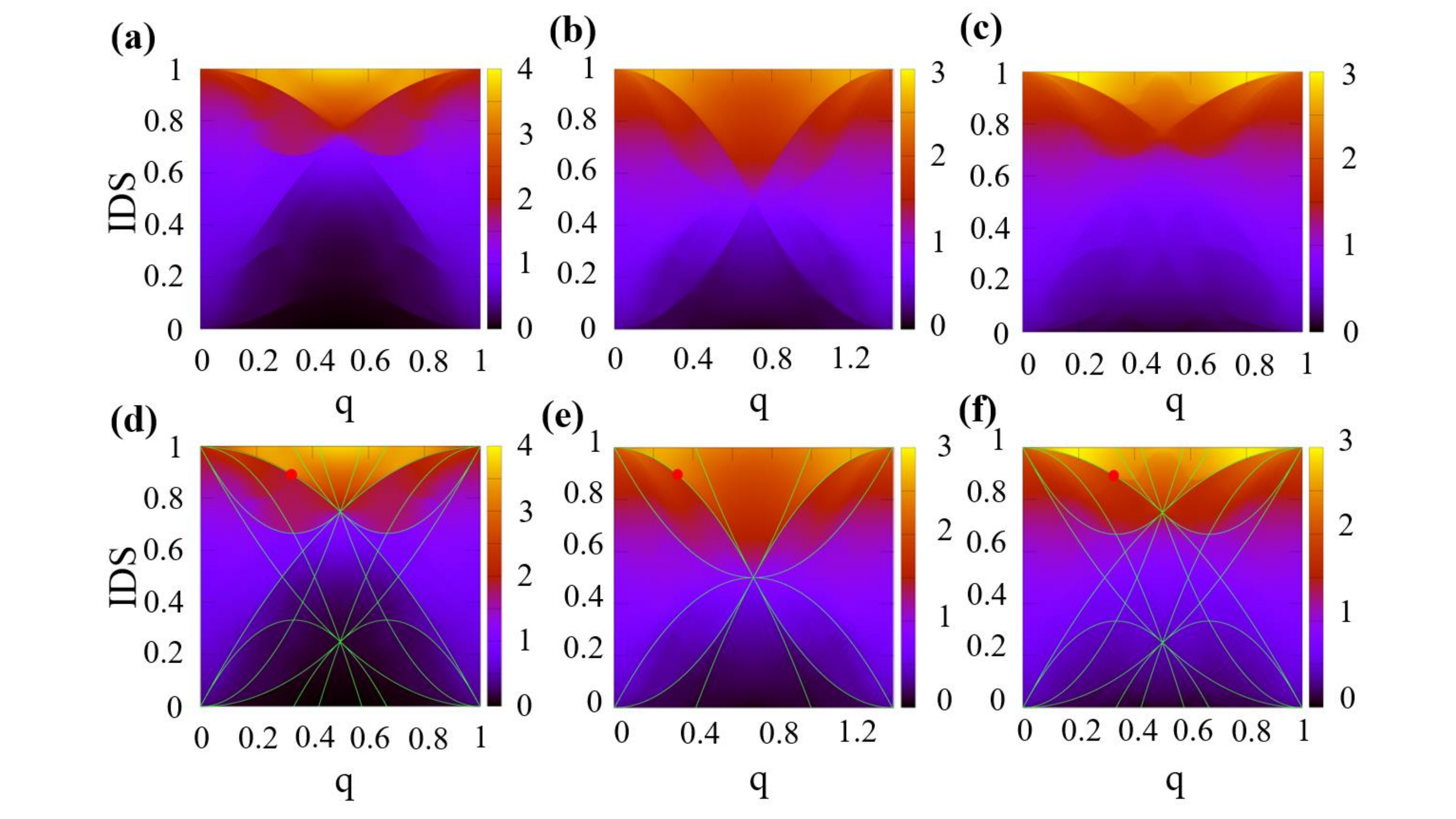}
\caption{\small  {\bf Fig 7 Visualizing the topological invariants.} {\bf a}, {\bf b}, {\bf c} Integrated density of states (IDS) as computed from the spectra in Fig.~\ref{Fig:SpectralB}(a,b,c), respectively. The IDS values inside the spectral gaps can be identified by the abrupt changes in color. {\bf d}, {\bf e}, {\bf f} Fittings of the IDS values inside the spectral gaps, seen in panels {\bf a}, {\bf b}, {\bf c},  with expression~\eqref{Eq:IDSFit}. Each curve is determined by the six topological invariants associated to each spectral gaps. The marked curves correspond to the marked gaps in Fig.~\ref{Fig:SpectralB} and they can be fitted with $1-\theta^2$, indicating $n_{\{1,2,3,4\}} =-1$, hence a $2^{\rm nd}$ Chern number $-1$.}
\label{Fig:IDSFit}
\end{figure*}

Since every function over a torus can be Fourier decomposed, all $T_f$'s are linear combinations and powers of $d'$ operators $T_j$ corresponding to the elementary functions
\begin{equation}
u_j(\bm \phi) = e^{\imath 2 \pi \phi_j}, \quad \imath = \sqrt{-1}, \quad j = 1,\ldots, d',
\end{equation}
where the torus $\TM^{d'} = \RM^d/\Ll'$ is parameterized as $\bm \phi = \sum \phi_j \bm a'_j$, $\phi_j \in [0,1]$. The conclusion is that any Galilean invariant $D_{\bm \phi}$ belongs to the algebra generated by $d'+d$ operators, which are the $T_j$'s mentioned above together with the elementary shifts $T_{d'+j}=S_{\bm e_j}$, corresponding to the generators $\bm e_j$ of $\ZM^{d}$. These operators obey the following commutation relations
\begin{equation}\label{Eq:CommRel}
T_i \, T_j = e^{\imath 2 \pi \theta_{ij}} \, T_j \, T_i 
\end{equation}
with the matrix $\Theta = \{\theta_{ij}\}$ fully determined by the two lattices $\Ll$ and $\Ll'$. Specifically, if $A$ is the transformation matrix, $\bm a_j = \sum_{j=1}^{d'} A_{ji} \bm a'_i$, $j =\overline{1,d}$, then
 \begin{equation}
 \theta_{ij} = - \theta_{ji} = A_{ij}, \quad \ i=\overline{1,d'}, \quad j = \overline{1,d},
 \end{equation}
 and zero for the rest of the indices. In particular, for the acoustic crystal from Fig.~\ref{Fig1}, we had $\bm a_i = \theta \bm a'_i$, hence $\theta_{13}=-\theta_{31}=\theta_{24} = - \theta_{42} = \theta$. This is also the case for the pattern {\bf c} from Fig.~\ref{Fig:Patterns}, but the $\Theta$-matrix for the pattern {\bf b} from Fig.~\ref{Fig:Patterns} contain more entries. As a consequence, the non-commutative 4-torus is not just a simple product of two non-commutative 2-tori, as it was the case in the previous experimental works.\cite{LohseNature2018,ZilberbergNature2018}

\section{Visualizing the topological invariants} 
In Fig.~\ref{Fig:SpectralB} we illustrate the bulk spectrum of the generic dynamical matrix 
\begin{equation}
D_\Pp = \sum\limits_{\bm x,\bm y \in \Pp} e^{-1.5|\bm x -\bm y|^2 }\, |\bm x\rangle \langle \bm y |,
\end{equation}
evaluated evaluated on the three patterns shown in Fig.~\ref{Fig:Patterns}. The fractal nature of the spectra is evident in Fig.~\ref{Fig:SpectralB}. In the \emph{Supplementary Information}, we demonstrate that model dynamical matrices can indeed reproduce the resonant spectrum of the crystal reported in Fig.~\ref{Fig1}. Here, just for illustrative purposes, we chose to work with a model that opens larger gaps. In Fig.~\ref{Fig:SpectralDome}, we report the resonant spectrum as computed with open boundary conditions for two shapes, a square and an octagon. Spectral domes can be identified in both cases.

The Chern numbers of the topological bulk gaps can be evaluated directly using existing numerical techniques developed for aperiodic systems.\cite{ProdanSpringer2017} However, due to the large number of gaps in Fig.~\ref{Fig:SpectralB} and to the large number of (strong and weak) topological invariants per gap, an alternative high-throughput method is needed. In fact, the method based on the K-theoretic gap labels\cite{Bellissard1986,Bellissard1995} explained below enables us to visualize the complete set of invariants associated with a gap. Besides the topological invariants, this method supplies additional predictions that can be tested against the numerical simulations and, as we shall see, this confirms beyond any doubt that the algebra of observables is indeed the non-commutative 4-torus.

In Fig.~\ref{Fig:IDSFit}(a-c), we report the integrated densities of states (IDS) as computed from the spectra Fig.~\ref{Fig:SpectralB}(a-c), respectively, using the usual definition
\begin{equation}\label{Eq:IDS1}
{\rm IDS}(E) = \frac{\# \ {\rm of} \ {\rm eigenvalues} \ {\rm below} \ E}{\# \ {\rm of} \ {\rm resonators}}.
\end{equation}
The IDS is plotted as function of $\theta$ and energy $E$, with the latter along the axis coming out of the paper. Since the view point for this graph is from above, we encode the values of the energy in a color map. The abrupt changes in color correspond to the cases when $E$ resides inside the bulk spectral gaps, because in that case the 3-dimensional graph of the IDS shoots straight out of the paper. The striking observation is that these features are not random. Perhaps even more striking is that same patterns are seen if one repeats the calculations with a different dynamical matrix. The reason behind these facts is that the features seen in the IDS plots are not determined by the dynamical matrix but rather by the K-theory of the algebra of observables. Indeed, if $P_G$ represents the gap projection for a gap $G$, then ${\rm IDS}(G)$ can be equivalently computed as the trace per area of $P_G$, ${\rm IDS}(G)=\Tt(P_G)$. K-theory for operator algebras\cite{BlackadarBook} classifies these projections up to stable homotopy and organizes them as an abelian group known as the $K_0$-group (see \emph{Supplementary Information}). For the non-commutative $s$-torus, the $K_0$-group is generated by $2^{s-1}$ projections $[e_J]$, where $J$ is a subset of indices drawn from $\{1,\ldots,s\}$ and the cardinal of $J$ is even. Now, the gap projection defines a class inside the $K_0$-group and we have the decomposition $[P_G]=\sum_{J} n_J [e_J]$ into the generators. The integer numbers $n_J$, known as the gap labels,\cite{Bellissard1986,Bellissard1995} represent the complete set of topological invariants that can be associated to a gap projection. They are related to the weak and strong Chern numbers (see \emph{Supplementary Information}), in particular, $n_{\{1,2,3,4\}}$ equals the $2^{\rm nd}$-Chern number. Our task is to extract the gap labels for the patterned acoustic crystals and that information is already contained in Fig.~\ref{Fig:IDSFit}. Indeed, 
\begin{equation}
{\rm IDS}(G)=\Tt(P_G) = \sum_J n_J \Tt(e_J),
\end{equation}
and, by using the value of the trace per area on the generators,\cite{Elliott1984} we obtain the following prediction for the features seen in Fig.~\ref{Fig:IDSFit}
\begin{equation}\label{Eq:IDSFit}
{\rm IDS}(G) = \sum_{J \subseteq \{1,\ldots,s\}} n_J \, {\rm Pfaffian}(\Theta_J), \quad |J|={\rm even},
\end{equation}
where $\Theta_J$ is the $\Theta$-matrix restricted to the set of indices $J$. As demonstrated in Figs.~\ref{Fig:IDSFit}(d,e,f), this expression (with $s=d+d'$) fits ALL the features seen in the IDS maps, despite the vast difference between the textures of the corresponding patterns (see Fig.~\ref{Fig:Patterns}). This serves as our proof that the dynamical matrices for the patterns in Fig.~\ref{Fig:Patterns} indeed belong to the non-commutative 4-torus and that we are witnessing the 4D quantum Hall physics.

Using the fittings from Fig.~\ref{Fig:IDSFit}, we were also able to extract the topological invariants associated to the gaps seen in the spectra from Fig.~\ref{Fig:SpectralB}. Furthermore, using more optimized discrete models (see \emph{Supplementary Information}), we were able to conclude that ${\rm Ch}_2=-1$ for the spectral gap analyzed experimentally. 

\section{Discussion}

We found that the boundary physics of aperiodic crystals emulating 4D IQHE is much more interesting and complex than previously believed. While the bulk-boundary correspondence for the virtual higher-dimensional systems is well understood, its manifestation in the lower physical dimensions is not. The phason engineering introduced by our work will be a very effective tool for this research because it supplies a high throughput of topological systems, which is absolutely needed for a systematic investigation of the boundary Weyl physics of these systems. The principles behind the emergence of IQHE in these systems are extremely general and robust, in particular, they do not require fine tuning, hence they can be easily implemented in laboratories or embedded in different applications.

As demonstrated in Fig.~\ref{Fig3}, the higher dimensional topological phases supply fundamentally ways of topological wave steering, whose possible applications remain to be discovered. Nevertheless, we already envision radically directions in mode steering, which can be useful for information processing. Indeed, the phason trajectory reported in Fig.~\ref{Fig3} is special in two respects: it has non-trivial topology and it occurs at constant frequency. As such, a coherent drive of the phason along that trajectory will not only steer the mode around the sample, as seen in Fig.~\ref{Fig3}, but will also generate temporal de-phasings that can be computed as Berry phases. In fact, the bulk modes can be also manipulated in a similar way, by driving the phason along topologically distinct loops inside the phason space. As already envisioned in \cite[]{BarlasPRL2020} such controlled temporal de-phasings could be used for certain forms of information processing.  

We believe that the principles revealed in this work exhaust the many ways one can engineer the phason spaces. They show that, in principle, there is not limit on how high in the virtual dimensions one can go. However, in practice, we expect that the actual laboratory designs to become increasingly challenging and the quality of the topological gaps to wear off as higher virtual dimensions are being conquered. Of course, the next in line is the 6D IQHE, which can be accessed with linear, planar or 3-dimensional meta-material structures. The latter will require a straightforward generalization of the algorithms used in the present work. Let us recall that the bulk-boundary correspondence principle was worked out in arbitrary dimension in \cite[]{ProdanSpringer2016} where one can find explicitly solved models as well as an explanation of quantized physical responses. 

\acknowledgements All authors acknowledge support from the W. M. Keck Foundation. E. P. acknowledges additional support from the National Science Foundation through the grant DMR-1823800.


\begin{thebibliography}{1}
\bibitem{Footnote1} The bulk-boundary was established for flat boundaries, hence here we are assuming that it holds also for a large disk-shaped sample.
\end{thebibliography}
\end{document}